%% file: main.tex
\documentclass[lettersize,journal]{IEEEtran}
\usepackage[utf8]{inputenc}
\usepackage{graphicx}
\usepackage{hyperref}
\hypersetup{colorlinks,
citecolor=black,
filecolor=black,
linkcolor=black,
urlcolor=black}
\usepackage{datetime}
\usepackage{amsmath}
\usepackage{float}
\usepackage{subcaption}
\usepackage{pdfpages}
\usepackage[english]{babel}
\newdateformat{monthdayyeardate}{%
  \monthname[\THEMONTH]~\THEDAY, \THEYEAR}
\usepackage{fancyhdr}
\usepackage{amsmath,amssymb,amsfonts}
\usepackage{graphicx}
\usepackage{textcomp}
\usepackage{gensymb}
\usepackage{tablefootnote}
\usepackage{algpseudocode}
\usepackage{algorithm}
\usepackage{xcolor}
\usepackage{float}
\usepackage{tikz}
\usetikzlibrary{shapes , arrows , positioning}
\graphicspath{{./}}
\usepackage[nottoc]{tocbibind}

\newcommand{\srcs}{SRC$_s$ }
\newcommand\norm[1]{\left\lVert#1\right\rVert}

\usepackage{cite}
\begin{document}
	\input{node_c.tex}
	\bibliography{refs.bib}
	\bibliographystyle{IEEEtran}
\end{document}

%% file: node_c.tex
\title{Node Cardinality Estimation in the Internet of Things Using Privileged Feature Distillation}
\author{Pranav S. Page, Anand S. Siyote, Vivek S. Borkar, Gaurav S. Kasbekar  
\thanks{P.S. Page, V.S. Borkar, and G.S. Kasbekar are with the Department of Electrical Engineering, Indian Institute of Technology (IIT) Bombay, Mumbai, India. A.S. Siyote is with TIH Foundation for IoT and IoE, IIT Bombay, Mumbai, India. Their email addresses are pranavpage33@gmail.com,  borkar.vs@gmail.com, gskasbekar@ee.iitb.ac.in, and anand.siyote@tihiitb.org, respectively. The contributions of V.S. Borkar and G.S. Kasbekar have been supported by the grant RD/0222-EETIHBY-014. }
}
\maketitle

\begin{abstract}
   The Internet of Things (IoT) is emerging as a critical technology to connect resource-constrained devices such as sensors and actuators as well as appliances to the Internet. In this paper, we propose a novel methodology for node cardinality estimation in wireless networks such as the IoT and Radio-Frequency IDentification (RFID) systems, which uses the privileged feature distillation (PFD) technique and works using a neural network with a teacher-student model. The teacher is trained using both privileged and regular features, and the student is trained with predictions from the teacher and regular features. We propose node cardinality estimation algorithms based on the PFD technique for homogeneous as well as heterogeneous wireless networks.  We show via extensive simulations that the proposed PFD based algorithms for homogeneous as well as heterogeneous networks achieve much lower mean squared errors in the computed node cardinality estimates than state-of-the-art protocols proposed in prior work, while taking the same number of time slots for executing the node cardinality estimation process as the latter protocols. 
\end{abstract}


\section{Introduction}
The Internet of Things (IoT) is emerging as a critical technology to connect a large number of resource-constrained devices such as sensors and actuators as well as appliances to the Internet \cite{wu2011m2m}. 
Many industries, including smart grids, healthcare, vehicular telematics, smart cities, security and public safety, agriculture, and industrial automation, extensively use IoT networks \cite{liu2014design}. Active research is being conducted on designing effective networking protocols to handle the growing number of IoT devices.
The design of medium access control (MAC) protocols for IoT networks is particularly challenging because of their unique characteristics \cite{rajandekar2015survey}. For instance, (i) network access must be provided to a large number of IoT devices, (ii) most IoT devices are battery-powered and have limited power availability, and (iii) the quality of service (QoS) requirements in IoT applications differ from those in human-to-human (H2H) communications \cite{rajandekar2015survey}. One of the key components of a MAC protocol for IoT networks is a node cardinality estimation protocol that rapidly estimates the number of active devices (i.e., the devices that currently have some data that needs to be transferred to the base station) in every time frame \cite{rajandekar2015survey}. These estimates can be used to determine the optimal values of various MAC protocol parameters such as contention probability, contention period duration, data transmission period duration, etc. \cite{kadam2017fast, kadam2020fast, liew2019probability}. 

Node cardinality estimation protocols also have a large number of applications apart from their use in MAC protocol design. They are used in \cite{kanistras2013survey} to periodically estimate the numbers of vehicles moving on various congested routes; the estimated information can be used to dynamically adapt the ON/ OFF periods of traffic lights based on vehicle density. Consider a farm with sensors installed to track a number of variables such as temperature and soil moisture. Before gathering the actual data from the active sensors, a mobile base station (MBS), e.g., one mounted on an unmanned aerial vehicle (UAV), navigates over the agricultural area and stops at designated locations to estimate the number of active sensors \cite{giambene20195g}. This increases the effectiveness of data collection because the MBS can optimally determine the amount of time it needs to spend at each stop when it subsequently returns to the same locations to collect the actual data and because it can inform the active sensors when to be available for sending the data based on the estimates. During or after natural calamities such as floods and earthquakes, MBSs hover above the affected area to estimate the number of people who need help. These estimates are used to plan disaster relief efforts and efficiently distribute supplies \cite{dinh2019flying}. Also, numerous Radio-Frequency IDentification (RFID) systems use node cardinality estimation protocols for inventory management, tag identification, missing tag detection, etc. \cite{arjona2018timing, liu2019novel, liu2020time, yu2018efficient, yu2018fast, liu2022revisiting, fahim2018towards, zhu2019collisions, zhang2018missing, liu2020retwork}.

Extensive research has been conducted on the problem of node cardinality estimation in IoT networks and RFID systems. Most of this research is focused on node cardinality estimation in a \emph{homogeneous} network, wherein the network consists of only one type of nodes \cite{qian2011cardinality,arjona2017scalable, hou2015place, liu2019collision, lin2019tash, zhou2018rfid, ng2020fast, duan2016d,ashrafuzzaman2018efficient,oh2015joint,liu2017novel,tavana2018congestion,el2018machine,bui2017novel,lin2015estimation,shirvanimoghaddam2016massive, kodialam2007anonymous,zheng2011pet,zheng2013zoe,gong2014arbitrarily,liu2016fast,liu2016rfid}. In addition, some work has been carried out on node cardinality estimation in  a \emph{heterogeneous} network, that is,  a network consisting of $T$ types of nodes, where $T \geq 2$ is an integer \cite{prasad2019rapid, kadam2021rapid, kadam2020node, kadam2023node, liu2014design, kadam2017fast, kadam2020fast, xiao2019estimating, zhang2020efficient, liu2016multi}. Different types of nodes in a heterogeneous network may have different hardware and software capabilities such as different processor speeds, transmission power, memory, battery life, etc.; also, different types of nodes may have different QoS requirements \cite{kadam2021rapid}. 
In a heterogeneous network with  $T$ types of nodes, execution of an estimation protocol designed for a homogeneous network $T$ times to obtain separate estimates of the cardinality of each node type is inefficient and improved protocols that rapidly obtain these estimates have been proposed in \cite{prasad2019rapid, kadam2021rapid, kadam2020node, kadam2023node, liu2014design, kadam2017fast, kadam2020fast, xiao2019estimating, zhang2020efficient, liu2016multi}. However, all the protocols designed so far for node cardinality estimation in homogeneous and heterogeneous networks use simple techniques, which are sub-optimal, for computing the node cardinality estimates.  Very little research has been conducted so far on applying machine learning based techniques to compute node cardinality estimates; in particular, to the best of our knowledge, the powerful \emph{Privileged Feature Distillation} (PFD) technique \cite{yang2022toward} has not been used in prior work for node cardinality estimation in wireless networks. This is the space in which we contribute in this paper. 

In this paper, we propose a novel methodology for node cardinality estimation in wireless networks such as the IoT and RFID sytems, which uses the PFD technique and works using a neural network with a \emph{teacher-student model} \cite{yang2022toward}. The teacher is trained using both privileged and regular features, and the student is trained using predictions from the teacher and regular features \cite{yang2022toward}. 

The concept of a privileged feature \cite{yang2022toward} arises in scenarios where a particular feature $z$ is available during the training phase but not during the testing or inference phase. The term ``privileged''  refers to the notion that this feature possesses additional information during the training process that can potentially aid in improving the prediction accuracy or performance.
By identifying privileged features and incorporating them into the training process, it is possible to leverage the additional information they provide to potentially improve the model's predictive capabilities, when those features are not available during the testing phase.
\emph{Distillation} \cite{hinton2015distilling} refers to the standard practice of labeling the training dataset using teacher predictions, and using these as supervision targets in the training of the student model. PFD has been successfully applied in various machine learning problems including speech recognition \cite{markov2016robust}, medical imaging \cite{gao2019privileged}, and image super-resolution \cite{lee2020learning}. We review some background concepts pertaining to PFD in Section \ref{bg:PFD}. 

The main contributions of this paper are as follows. In Section \ref{SSC:homogeneous:model:pf}, we formulate the problem of estimating the number of active nodes in a homogeneous wireless network, while minimizing the mean squared error (MSE) between the actual number of active nodes and the algorithm's estimate. In Section \ref{SSC:heterogeneous:model:pf}, we generalize this problem formulation to the problem of estimating the number of active nodes of each type in a heterogeneous wireless network with $T$ types of nodes, where $T \geq 2$ is an integer, while minimizing the MSE. In Section \ref{SSC:algorithms:homogeneous:network}, we propose a novel algorithm, which uses the PFD technique and a neural network with a teacher-student model, for node cardinality estimation in a homogeneous network. In Section \ref{SSC:algorithms:heterogeneous:network}, we generalize this algorithm for estimating the cardinality of each node type in a heterogeneous network with $T$ types of nodes. In Section \ref{SC:simulations}, we show via extensive simulations that the proposed PFD based algorithm for a homogeneous (respectively, heterogeneous) network achieves a much lower MSE than the state-of-the-art simple RFID counting (SRC$_s$) protocol \cite{chen2013understanding} (respectively, $T$-\srcs protocol), even though both the algorithms take the same number of time slots for executing the node cardinality estimation process.    

The rest of this paper is organized as follows. A review of related prior literature is provided in Section \ref{SC:related:work}. The system model and problem formulation are described in Section \ref{SC:system:model:problem:formulation}. Some relevant background is given in Section \ref{SC:background}. The proposed algorithms and other algorithms for comparison are described in Section \ref{SC:algorithms}. Simulation results are provided in Section \ref{SC:simulations}. Finally, conclusions and directions for future research are provided in Section \ref{SC:conclusions:future:work}.

\section{Related Work}
\label{SC:related:work}
In Section \ref{SSC:related:work:node:cardinality:estimation} (respectively, Section \ref{SSC:related:work:PFD}), we provide a review of related prior literature on node cardinality estimation in wireless networks (respectively, PFD). 
\subsection{Node Cardinality Estimation}
\label{SSC:related:work:node:cardinality:estimation}
The estimation of active node cardinalities is considered crucial in the design of a MAC protocol for IoT networks. This importance has led to extensive research being conducted on this topic \cite{ng2020fast,duan2016d,ashrafuzzaman2018efficient,oh2015joint,liu2017novel,tavana2018congestion,el2018machine,bui2017novel,lin2015estimation,shirvanimoghaddam2016massive}. These studies focus not only on estimating the number of active devices in a homogeneous IoT network, but also on using these estimates to determine the contention probabilities that optimize the throughput of their respective MAC protocols for IoT networks. To estimate the number of active nodes in the current time frame, the estimation scheme proposed in  \cite{duan2016d} uses the estimates obtained in the previous frame as well as the sub-optimal Dynamic Access Class Barring (D-ACB) factors from the previous frame. In \cite{ashrafuzzaman2018efficient},  a modified Carrier Sense Multiple Access/ Collision Avoidance (CSMA/CA) protocol intended for IoT networks was introduced. The size of the backoff window for the current time frame is chosen by this protocol by considering the size of the previous backoff window and the previously computed estimates of the active node cardinality. This procedure incorporates historical estimates to improve the effectiveness of the backoff mechanism.
Note that both  \cite{duan2016d} and  \cite{ashrafuzzaman2018efficient} relied on the estimates obtained in previous frames to compute their estimates in the current frame. This iterative approach allows the utilization of past information to improve the accuracy and effectiveness of the estimation process.
A new technique for dynamic access control and resource allocation for random-access channels based on an estimation scheme was introduced in  \cite{oh2015joint}. The only input used in the estimation procedure in  \cite{oh2015joint} for computing the estimates was the number of open slots.
The 6-Dimensional Markov Chain (6-DMC) estimation method was introduced in  \cite{liu2017novel}. The numbers of devices that are delay tolerant (DTDs) and delay sensitive (DSDs) are estimated using this approach. The estimation methods in  \cite{liu2017novel},  \cite{tavana2018congestion}, and  \cite{el2018machine} are based on 6-DMC, Maximum Likelihood Estimation (MLE), and IoT-OSA (an extension of the opportunistic splitting technique).

The satellite random access (RA) MAC protocol is provided in  \cite{bacco2018tcp}. In this protocol, throughput is maximized by computing an estimate of the number of Return Channel Satellite Terminals (RCSTs). The number of collisions observed in earlier frames affects the length of the current frame in the model described in \cite{bacco2018tcp}. The approach described in  \cite{bui2017novel} estimates the number of nodes that cause collisions. In Long-Term Evolution (LTE) networks, this estimation enables an effective partitioning of nodes into a predetermined number of groups while minimizing intra-group collisions. Dynamic Backoff (DB), a new method for resolving channel contention, was first described in  \cite{lin2015estimation}. Based on the estimated number of competing active devices, this approach modifies the size of the backoff window used to manage channel contention during data transfer. The scheme proposed in  \cite{lin2015estimation} also dynamically modifies each frame size based on the projected number of devices, making it adaptable to shifting network circumstances and device activity.

The node cardinality estimation problem in IoT networks is similar to the tag cardinality estimation problem in the context of RFID technology. In the latter situation, an RFID reader estimates the number of tags, like a base station does when estimating the number of active nodes in an IoT network. Schemes for estimating the number of tags in an RFID system were proposed in  \cite{arjona2017scalable, hou2015place, liu2019collision, lin2019tash,chen2013understanding,qian2011cardinality,kodialam2007anonymous,zheng2011pet,zheng2013zoe,gong2014arbitrarily,zhou2018rfid,liu2016fast,liu2016rfid}.

Node cardinality estimation schemes for heterogeneous IoT networks and RFID systems have been proposed in \cite{liu2014design, kadam2017fast, kadam2020fast, xiao2019estimating, zhang2020efficient, liu2016multi, kadam2021rapid, prasad2019rapid, kadam2020node, kadam2023node}. A specialized MAC protocol for a heterogeneous IoT network, catering to three types of IoT devices, has been introduced in \cite{kadam2017fast, kadam2020fast}. It incorporates a rapid estimation protocol to determine active node counts, and uses them to optimize the contention probabilities in the MAC protocol. An efficient node cardinality estimation solution with two components-- snapshot collection and accurate estimation-- has been given in \cite{xiao2019estimating}. It focuses on improving joint cardinality estimation in distributed RFID systems, allowing queries across multiple tag sets at different locations and times with controlled error. It has applications in tracking product flows in logistics. Simulations show a significant time cost reduction while maintaining accuracy.  Enhancement of  RFID technology's cardinality estimation function in two ways-- joint estimation across tags at different locations and times and category-level tracking-- was proposed in \cite{zhang2020efficient}. It introduces an anonymous protocol that efficiently estimates joint category-level information, preserving tag anonymity and enabling applications such as monitoring diverse products in distributed supply chains. Multi-category RFID tag estimation has been addressed in \cite{liu2016multi}, aiming to swiftly and accurately count tags within each category. It introduces the ``Simultaneous Estimation for Multi-category RFID Systems'' (SEM) approach, leveraging Manchester coding to decode combined signals, allowing simultaneous estimation across categories while maintaining pre-defined accuracy. SEM significantly improves estimation speed compared to existing protocols. Rapid estimation of the cardinalities of active nodes of different types in heterogeneous IoT networks with $T$ node types, where $T \geq 2$ is an arbitrary integer, has been addressed in \cite{prasad2019rapid, kadam2021rapid, kadam2020node, kadam2023node}. 

However, the PFD technique has not been applied to the problem of node cardinality estimation in either a homogeneous or heterogeneous wireless network in any of the above prior works. 

\subsection{Privileged Feature Distillation (PFD)} 
\label{SSC:related:work:PFD}
The concept of learning with privileged features was introduced in \cite{vapnik2009new}, and a framework called ``learning using privileged information'' (LUPI) was proposed. Privileged information is the primary approach used by LUPI to discriminate between simple and complex cases. These methods are closely related to Support Vector Machines (SVM); e.g., the ``SVM+''  algorithm, which creates slack variables from privileged features and learns an SVM based on regular features with those slack variables, is proposed in \cite{vapnik2009new, pechyony2010smo}. A pair-wise SVM algorithm for ranking that uses privileged features to differentiate between easy and hard pairs is proposed in \cite{sharmanska2013learning}. The privileged features are employed in the version presented in \cite{lapin2014learning} to produce importance weighting for various training samples.

A popular technique for knowledge transfer is model distillation  \cite{hinton2015distilling}, often from a large model to a smaller model  \cite{polino2018model,gou2021knowledge}. Recent studies  \cite{tang2018ranking,hofstatter2020improving,reddi2021rankdistil}, and even those where the teacher model and student model have the same structure  \cite{furlanello2018born,qin2021born}, have demonstrated remarkable empirical success in ranking tasks.

``Generalised distillation'' (GenD) is the term for the method first suggested in \cite{lopez2015unifying} for using distillation to learn from privileged features. This offers a comprehensive perspective on distillation and LUPI. GenD and its derivatives  \cite{markov2016robust,garcia2019learning,lee2020learning} train an expert model using just privileged features, after which the student model is trained to replicate the expert's predictions. Recently, PFD was presented in  \cite{xu2020privileged}, where the teacher model accepts input from both regular and privileged features. Due to their emphasis on privileged feature exploitation rather than model size reduction, PFD and GenD are different from traditional model distillation. On a non-public data collection, the improved performance of PFD for recommendation systems is empirically demonstrated in \cite{xu2020privileged}.

Despite the aforementioned empirical accomplishment, there remains a lack of understanding of privileged characteristics distillation. Prior research  \cite{pechyony2010theory} demonstrates that LUPI speeds up convergence under the strict premise that the best classifier can be realised using just privileged information. GenD has a quick convergence rate, as shown by \cite{lopez2015unifying}. This is different from PFD since it assumes that the function class complexity of the student model is significantly larger than that of the teacher model. The study by \cite{gong2018teaching} on GenD under semi-supervised learning reveals that the benefits come from reducing the complexity of student function classes. However, it does not quantify this reduction, and the theory does not explain why exploiting privileged traits is advantageous. 

Prior proposals include other uses of privileged features. To enhance picture classification performance,  \cite{chen2017training} learns a more varied representation using privileged information.
Distillation strategies have been proposed by  \cite{lee2020learning,wang2021privileged} for more effective feature extraction from regular features.
A more recent study  \cite{collier2022transfer} examined the possibility of training a model using both regular and privileged features to improve the internal representation of regular features.

However, to the best of our knowledge, this paper is the first to use the technique of PFD to address the problem of node cardinality estimation in wireless networks.

\section{System Model and Problem Formulation}
\label{SC:system:model:problem:formulation}
In Section \ref{SSC:homogeneous:model:pf} (respectively, Section \ref{SSC:heterogeneous:model:pf}), we describe the system model and problem formulation for the case of a homogeneous network (respectively, heterogeneous network). 

\subsection{Homogeneous Network}
\label{SSC:homogeneous:model:pf}
\subsubsection{System Model}
Consider a population of nodes such that each node is in the range of a single stationary base station (BS). We consider a node as \emph{active} when it has data to send to the BS. Time is divided into frames of equal durations.
To effectively design MAC protocols for data upload to the BS, the number of active nodes in a time frame must be estimated. We study the case, which often arises in practice, when there exists some correlation between the number of nodes active in a time frame and the number of nodes active in the next time frame. E.g., the number of active nodes may evolve as a discrete-time Markov chain. 

\subsubsection{Problem Formulation}
The aim of this work is to design algorithms that minimize the mean squared error (MSE) between the actual number of active nodes in a time frame and the algorithm's estimate of this number, while also reducing the number of time slots needed to produce the estimate. The objective of minimizing the MSE for a homogeneous network of nodes is as follows:
\begin{equation}
\label{eq:homo_problem_eq}
	alg' = \mbox{argmin}_{alg} \mathbb{E} \left[  \lim_{\tau \to \infty}\frac{1}{\tau}\sum_{t=0}^{\tau - 1}(\hat{n}^{alg}_t - n^{truth}_t)^2 \right].
\end{equation}
Here, $n^{truth}_t$ is the true number of nodes active in time frame $t$, while $\hat{n}^{alg}_t$ is the estimate given by the algorithm $alg$ in time frame $t$. The expectation is with respect to the different realizations of the random process ($n^{truth}_t, \, t = 0,1,2,\ldots$). The objective is to find the algorithm $alg'$ that achieves the minimum in the RHS of (\ref{eq:homo_problem_eq}).

\subsection{Heterogeneous Network}
\label{SSC:heterogeneous:model:pf}
\subsubsection{System Model}
\begin{figure}
    \centering
    \includegraphics[width=0.3\textwidth]{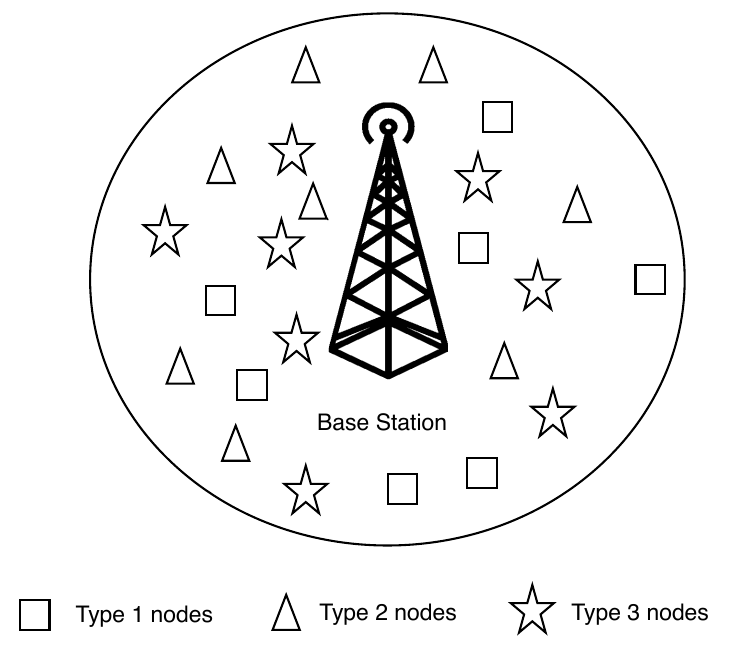}
    \caption{The figure shows an example heterogeneous network with $T=3$ types of nodes in the range of a BS.}
    \label{fig:network_model}
\end{figure}
In the heterogeneous case, there exist $T$ types of nodes, where $T \geq 2$ is an integer, in the range of a BS, as shown in Figure \ref{fig:network_model} for the case $T=3$. For example, nodes of different types may correspond to nodes that send emergency traffic such as fire alarms, nodes that contain moisture sensors, nodes that contain temperature sensors, etc. We represent the numbers of active nodes of different types by $\mathbf{n}^{truth}_t$, a $1\times T$ vector, where $\mathbf{n}^{truth}_t[b]$, $b \in \{1, \cdots, T\}$, is the number of active nodes of type $b$ in time frame $t$.
\subsubsection{Problem Formulation}
 Similar to the homogeneous case, for the heterogeneous case, the objective is to minimize the expected time-averaged squared Euclidean distance between the estimates computed by the algorithm and the time series, $\mathbf{n}^{truth}_t$, of true numbers of active nodes:
 \begin{equation}
\label{eq:het_problem_eq}
	alg' = \mbox{argmin}_{alg} \mathbb{E} \left[  \lim_{\tau \to \infty}\frac{1}{\tau}\sum_{t=0}^{\tau - 1}\norm{\hat{\mathbf{n}}^{alg}_t - \mathbf{n}^{truth}_t}^2_2 \right].
\end{equation}
Here, $\hat{\mathbf{n}}^{alg}_t$ is a $1 \times T$ vector, where $\hat{\mathbf{n}}^{alg}_t[b]$, $b \in \{1, \cdots, T\}$, is the estimate of the number of active nodes of type $b$ in time frame $t$ found by the algorithm $alg$. The objective is to find the algorithm $alg'$ that achieves the minimum in the RHS of (\ref{eq:het_problem_eq}).

\section{Background}
\label{SC:background}
In this section, we review some concepts that are used in the rest of the paper. 
\subsection{Simple RFID Counting (SRC$_s$) Protocol}
\label{bg:srcs}
\begin{figure}
    \centering
    \includegraphics[width=0.5\textwidth]{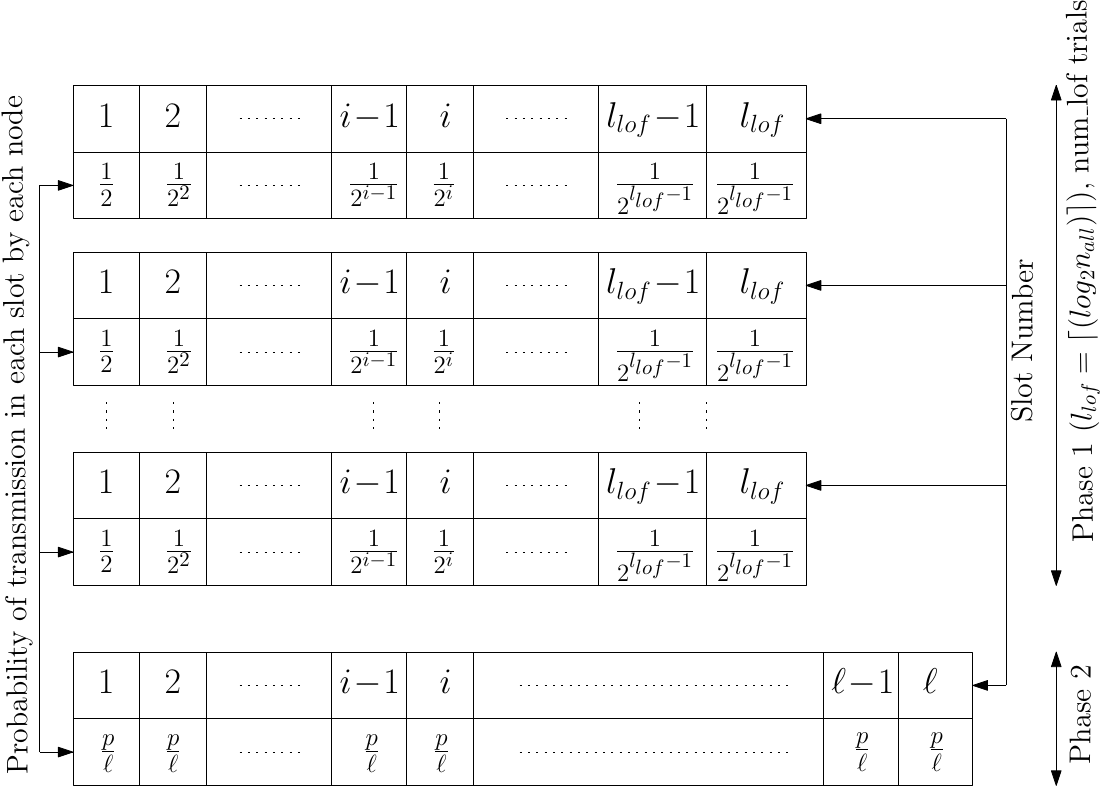}
    \caption{The figure shows the frame structure of the \srcs protocol.}
    \label{fig:frame_srcs}
\end{figure}

\srcs is a node cardinality estimation protocol for a homogeneous network \cite{chen2013understanding}, which finds an estimate, $\hat{n}$, of the number of active nodes, $n$, to within given accuracy requirements $\epsilon$ and $\delta$, i.e, the following relation is satisfied: 
\[\
\mathbb{P}(|\hat{n} - n| \leq \epsilon n) \geq 1 - \delta. 
\]
The \srcs protocol (Algorithm \ref{alg:srcs})  \cite{chen2013understanding} uses the Lottery Frame (LoF) protocol (Algorithm \ref{alg:lof}) to generate a rough estimate of the number of active nodes, followed by a Balls and Bins (BB) trial (Algorithm \ref{alg:bb}) that uses the rough estimate given by LoF. The LoF protocol uses a trial length of $l_{lof} = \lceil \log_2{n_{all}} \rceil$ time slots, where $n_{all}$ is the maximum number of active nodes in the network. The \srcs protocol conducts multiple, say num\_lof, LoF trials and computes an average of the rough estimates generated in the trials, $n'$. The choice of the length of the BB trial $l$ depends on the relative error tolerated, $\epsilon$, and is taken as $l=\frac{65}{(1-0.04^\epsilon)^2}$ \cite{chen2013understanding}. The number of LoF trials, num\_lof, in \srcs is taken to be of the order $O(\log{\frac{1}{\delta}})$. For example, for $\delta=10^{-3}$, num\_lof $= 3$ is used. 
The output of the BB trial is the final SRC$_s$ estimate, $\hat{n}$. The frame structure of \srcs is shown in Figure \ref{fig:frame_srcs}.
\begin{algorithm}
	\caption{Lottery Frame Protocol \cite{chen2013understanding}}
	\label{alg:lof}
	\begin{algorithmic}[1]
		\State Choose trial length $l_{lof} = \lceil \log_{2}{n_{all}} \rceil$
		\State Each active node independently transmits in slot $i = 1, \ldots, l_{lof} - 1$ with probability $2^{-i}$ and in slot $l_{lof}$ with probability $2^{-(l_{lof} - 1)}$
		\State Trial ends when first empty slot (slot in which no node transmits), say slot $j$, is seen
		\State \Return $n' = 1.2897 \times 2^j$
	\end{algorithmic}
\end{algorithm}
\begin{algorithm}
	\caption{Balls and Bins Protocol \cite{chen2013understanding}}
	\label{alg:bb}
	\begin{algorithmic}[1]
		\State Given rough estimate $n'$, each active node independently participates in the trial of length $l$ slots with probability $p = \min(1, 1.6l/n')$ 
		\State Each participating node transmits in a slot chosen uniformly at random from the $l$ slots
		\State $z$ = number of empty slots in trial
		\If{$z>0$}
		\State \Return $\frac{\ln{(z/l)}}{\ln{(1- p/l)}}$
		\Else 
		\State \Return arbitrary number
		\EndIf
	\end{algorithmic}
\end{algorithm}
\begin{algorithm}
	\caption{SRC$_s$ Protocol \cite{chen2013understanding}}
	\label{alg:srcs}
	\begin{algorithmic}[1]
		\State Conduct num\_lof LoF trials and compute the following rough estimate: $n' = 1.2897 \times 2^{\sum_{m=1}^{\mbox{num\_lof}} (j(m) - 1)/ \mbox{num\_lof}}$, where $j(m)$ is the first empty slot in the $m'$th  LoF trial
		\State Run a BB trial of length $l$ in which each node participates with probability $p = \min(1, 1.6l/n')$
		\State Count the number of empty slots,  say $z$, in the trial
		\If{$z>0$}
		\State \Return $\frac{\ln{(z/l)}}{\ln{(1- p/l)}}$
		\Else 
		\State \Return arbitrary number
		\EndIf
	\end{algorithmic}
\end{algorithm}
\subsection{3-Stage Scheme-Balls and Bins (3-SS-BB) Protocol}
\label{bg:3ss}
\begin{figure}
	\centering
	\includegraphics[width=0.40\textwidth]{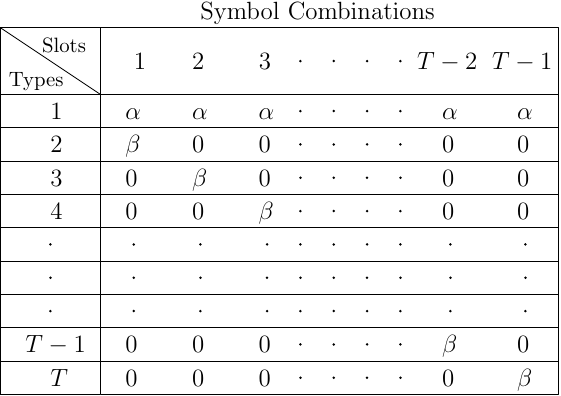}
	\caption{The figure shows the symbol combinations used by different types of nodes under the 3-SS-BB protocol. $0$ indicates no transmission.}
	\label{fig:3ss_pattern}
\end{figure}

3-SS-BB is a protocol for finding an estimate, $\hat{n}_b$, of the number of active nodes, $n_b$, of each type $b \in \{1, \ldots, T\}$ in a heterogeneous network with $T$ types of nodes (see Figure \ref{fig:network_model}) \cite{prasad2019rapid, kadam2021rapid}.    
It is an extension of the BB trial (Algorithm \ref{alg:bb}) to a heterogeneous network. It assumes that rough estimates, $n'_b$, $b \in \{1, \ldots, T\}$, of the numbers of active nodes of the $T$ types are initially available; e.g., these estimates may be generated by separately conducting LoF trials for each node type as in the first stage of SRC$_s$. 3-SS-BB uses a trial with $l$ blocks; within each block, there are $T-1$ slots. Each active node of type $b$ independently participates in the trial of $l$ blocks with probability $p_b = \min(1, 1.6l/n'_b)$. 
Each participating node of type $b$ chooses a block out of the $l$ blocks uniformly at random and sends, in its chosen block, the symbol combination of length $T-1$ slots given in the row corresponding to type $b$ in Figure \ref{fig:3ss_pattern}.  For example, each node of type 1 transmits the pattern $(\alpha, \alpha, \cdots, \alpha)$, while each node of type 2 transmits $(\beta, 0, \cdots, 0)$, where $\alpha$ and $\beta$ are distinct symbols (bit patterns) and $0$ indicates no transmission. If there are two or more transmissions in a slot, the result of the slot is $c$ (collision). Thus, a slot can have four possible outcomes: $\{0, \alpha, \beta, c\}$. Estimates $\hat{n}_b$, $b \in \{1, \ldots, T\}$, are generated based on the outcomes of the $(T-1)l$ slots using the algorithm provided in \cite{prasad2019rapid, kadam2021rapid}. 3-SS-BB is summarised in Algorithm \ref{alg:3_ss_bb}.
\begin{algorithm}
	\caption{3-SS-BB}
	\label{alg:3_ss_bb}
	\begin{algorithmic}[1]
        \State Given rough estimate $n'_b$, each active node of type $b$ independently participates in the trial of $l$ blocks with probability $p_b = \min(1, 1.6l/n'_b)$ 
		\State Each participating node chooses a block out of the $l$ blocks of the trial uniformly at random
		\State Each node of type $b$ sends, in its chosen block, the symbol combination of length $T-1$ slots given in the row corresponding to type $b$ in Figure \ref{fig:3ss_pattern} 
		\State In each of the resultant $(T-1)l$ slots, the outcome is one of $\{0, \alpha, \beta, c\}$
        \State Estimates $\hat{n}_b$, $b \in \{1, \ldots, T\}$, are generated based on the outcomes of the $(T-1)l$ slots
	\end{algorithmic}
\end{algorithm}
\subsection{Privileged Features Distillation (PFD)}
\label{bg:PFD}
In some problem settings, there exist some features that are not available during testing, but are available offline for training. Instead of discarding these features, one approach is to train a `teacher' model on the privileged features, say $\mathbf{x}_{privileged}$ \cite{yang2022toward}. The teacher model is then used in the training of a different `student' model \cite{yang2022toward}. The student is trained only on the features, say $\mathbf{x}_{general}$, that are available during testing, but the loss of the model is designed (see (\ref{eq:distillation})) as a convex combination of the student loss and the teacher loss. In  (\ref{eq:teacher_loss_math}), $\mathcal{L}^{teacher}$ refers to the teacher loss corresponding to the loss described by the loss function $L(\cdot, \cdot)$, operating on the prediction of the teacher and the target $y$. The teacher network is represented by $g^{teacher}$,  and the teacher's prediction is $g^{teacher}(\mathbf{x}_{privileged})$. Similarly, in (\ref{eq:student_loss_math}), the data loss $\mathcal{L}^{data}$ is the loss between the student's prediction $g^{student}(\mathbf{x}_{general})$ on the general features, $\mathbf{x}_{general}$, and the target $y$.
The mixing ratio between the data loss and the teacher loss is $\alpha \in [0,1]$. 
The student does not interact with the privileged features, nor with the teacher's predictions, but only with the loss between the teacher's prediction and the target. 
\begin{align}
\label{eq:teacher_loss_math}
	\mathcal{L}^{teacher} &= L(g^{teacher}(\mathbf{x}_{privileged}), y) \\
\label{eq:student_loss_math}
	\mathcal{L}^{data} &= L(g^{student}(\mathbf{x}_{general}), y)\\
\label{eq:distillation}
	\mathcal{L}^{student} &= \alpha \mathcal{L}^{data} + (1-\alpha) \mathcal{L}^{teacher}
\end{align}

\section{Algorithms}
\label{SC:algorithms}
In Section \ref{SSC:algorithms:homogeneous:network} (respectively, \ref{SSC:algorithms:heterogeneous:network}), we describe our proposed PFD based algorithm, and other algorithms for comparison, for homogeneous (respectively, heterogeneous) wireless networks. 

\subsection{Homogeneous Network}
\label{SSC:algorithms:homogeneous:network}
\subsubsection{Proposed Algorithm}
\label{subsubsec:homogeneous_proposed}
Consider the model and problem formulation described in Section \ref{SSC:homogeneous:model:pf}. The entire population  of nodes in the range of the base station is of a single type. Recall from Section \ref{bg:srcs} that the $\text{SRC}_s$ protocol consists of a LoF-based phase 1 and a BB-based phase 2. The LoF phase obtains a rough estimate of the number of nodes, $n'$, which the BB phase uses to obtain a refined estimate. 

In each time frame $t$, the proposed neural network (NN) based algorithm (see Algorithm \ref{alg:homo_nn}) executes only phase 2 (BB) and obtains the trial result. If the trial consists of $l_{BB}$ slots, then a vector of size $l_{BB}$ is generated via BB. This vector consists of the outcome (no transmission, success (one transmission), or collision) in each of the $l_{BB}$ slots of the BB trial. The trained model takes this vector as input, along with the estimate of the number of active nodes generated by the model in the previous time frame, and estimates the value of the number of active nodes in the current time frame. The NN is a student model trained using PFD as explained in Section \ref{subsubsec:homogeneous_training}; so henceforth, the trained model will be represented by the notation $Stu$. 

In time frame $0$, the proposed NN method conducts a set of LoF trials to give the initial rough estimate $\hat{n}'_0$, which is then used as the rough estimate (see step 1 in Algorithm \ref{alg:bb}) in a BB trial. The NN then uses the result of the BB trial, which is a vector of length $l_{BB}$, say $v_0$, and the rough estimate $\hat{n}'_0$, to generate its own estimate in time frame $0$, $\hat{n}_0^{Stu}$, using the student network $g^{Stu}$. Subsequently, in each  time frame $t = 1, 2, \ldots, \text{num\_iters}$, where $\text{num\_iters}$ is the total number of time frames, a BB trial is conducted with rough estimate $\hat{n}^{Stu}_{t-1}$ ($Stu$'s estimate of the previous time frame), which generates a vector of length $l_{BB}$, say $v_t$, as a result. The NN $g^{Stu}$ operates on $(v_t, \hat{n}^{Stu}_{t-1})$ to produce its estimate $\hat{n}^{Stu}_{t}$ in time frame $t$.

The motivation for using the estimate of the previous time frame, $\hat{n}^{Stu}_{t-1}$, as the rough estimate for the BB trial of the current time frame $t$ arises from the fact that some correlation exists between the nodes active in the previous time frame and the nodes active in the current time frame. We exploit this fact to reduce the number of time slots used by \emph{not} executing LoF trials to obtain the rough estimates for the BB trials of time frames $t = 1, 2, 3, \ldots$. 

\begin{algorithm}
	\caption{Proposed NN Method}
	\label{alg:homo_nn}
	\begin{algorithmic}[1]
		\State At $t=0$, conduct LoF trials to give the initial rough estimate $\hat{n}'_0$; then conduct BB trial to generate $v_0$
		\State $\hat{n}^{Stu}_0 = g^{Stu}(v_0, \hat{n}'_{0})$ 
		\For{$t=1, \cdots, \text{num\_iters}$}
			\State Conduct BB trial with rough estimate $\hat{n}^{Stu}_{t-1}$, giving result $v_t$
			\State  $\hat{n}^{Stu}_{t} = g^{Stu}(v_t, \hat{n}^{Stu}_{t-1})$
		\EndFor
	\end{algorithmic}
\end{algorithm}

\begin{figure}
	\centering
	\includegraphics[width=0.50\textwidth]{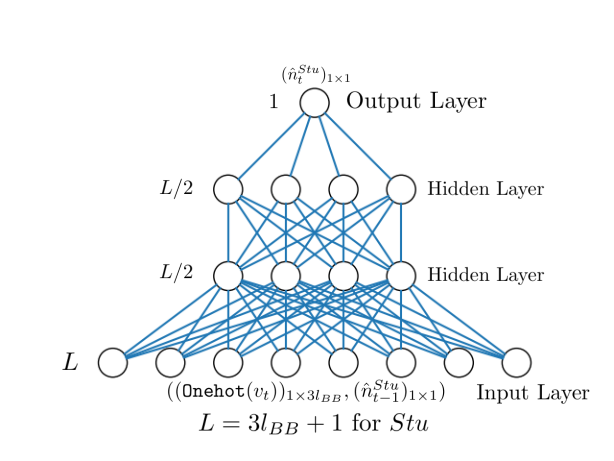}
	\caption{The figure shows the neural network architecture of a student ($Stu$) model used in the case of a homogeneous network.}
	\label{fig:homo_net}
\end{figure}

The architecture of the NN used is shown in Figure \ref{fig:homo_net}. The input dense layer of length $L$ has Rectified Linear Unit (ReLU) activation, while the other two $L/2$ dense layers (hidden layers) have sigmoid activation. The output layer of length $1$ has linear activation. A description of the activation functions is provided in \cite{dubey2022activation}.  The architecture has been designed for regression specifically and for ease of training, according to insights from \cite{hagan1997neural}. 

\subsubsection{Training}
\label{subsubsec:homogeneous_training}
The estimation of the number of active nodes in each time frame is to be done by a model that uses data recorded in a real online scenario. This implies that each element of the results of the trial $v_t$ would comprise of three possibilities: \{no transmission, success, collision\}. 
This would be a difficult problem to tackle, and would need more information than the online model has to perform well. Thus, we use PFD, where a teacher model is trained on privileged data not available in the online scenario, in particular, the number of nodes transmitting in each slot of the BB trial and the true value of the number of active nodes in the previous time frame. A student model is trained on the actual data seen during the online scenario. 

In particular, the objective is to train a student model, which is the actual $Stu$ NN used in Algorithm \ref{alg:homo_nn}, to perform online estimation of the number of active nodes, $\hat{n}^{Stu}_{t}$, in each time frame $t$ using the general feature vector $v_t$ and the previous time frame's estimate $\hat{n}^{Stu}_{t-1}$. The feature vector $v_t$ for a BB trial of length $l_{BB}$ has dimensions $1\times 3l_{BB}$ since the outcome of each of the $l_{BB}$ slots is represented using \emph{one-hot encoding} and there are 3 possibilities for a slot-- no transmission, success, and collision. Also, $L=3l_{BB}+1$ in Figure \ref{fig:homo_net} since the NN takes as input the vector $v_t$ and the estimate of the previous time frame. 

In the first step, the teacher model is to be trained. The teacher model is fed with privileged information. Let $V_t$ be a $1 \times l_{BB}$ vector denoting the results of the BB trial in time frame $t$ and containing privileged information. In particular, for $i \in \{1, \ldots, l_{BB}\}$, $V_t[i]$ is the number of active nodes transmitting in slot $i$ of the BB trial, which is privileged information since when a collision occurs, i.e., two or more nodes transmit in a slot, the number of transmitting nodes is not known in practice. 

Due to the nature of the algorithm, where the NN's output in the previous time frame is part of the input vector for the current time frame, if a network is trained on each time frame one-by-one, it overfits the current sample. It is thus able to predict the next few steps accurately, but the same model will `forget' the samples seen a few time frames ago. Hence, the procedure of getting a model to be a genie to `track' the evolution of the time series, logging the dataset, shuffling it, and training a new model to mimic the genie is followed. The model trained offline is able to go through the data multiple times and out-of-order. 

This calls for a step of data generation-- creating a dataset with $((V_t, n^{truth}_{t-1}), n^{truth}_{t})$ as the (feature, target) tuple for time frame $t$. The dataset is then shuffled and a new uninitialized teacher network is trained on this dataset.  

For data generation, recall that the BB trial requires a rough estimate of the number of active nodes. We use a \textit{genie} network for this purpose, as explained in Algorithm \ref{alg:teacher}. The \textit{genie} teacher is trained on the BB trial in each time frame $t$ after generating the prediction  $\hat{n}^{gTr}_t$, which is used as the initial rough estimate for the BB trial in time frame $t+1$. 

The \textit{genie} teacher network is represented by $g^{gTr}$. The estimate by the \textit{genie} teacher for the trial at time $t$ is $\hat{n}^{gTr}_t$, and is given by the following equation:
\begin{equation}
\label{eq:tr_prediction}
\hat{n}^{gTr}_t = g^{gTr}(V_t, n^{truth}_{t-1}).
\end{equation}
The teacher training is explained in Algorithm \ref{alg:teacher}.
In \textsc{gen\_teacher\_training\_data}, a new network $gTr$ is initialized, LoF trials are conducted to give an initial rough estimate $n'_0$, and a BB trial is conducted with rough estimate $n'_0$. The resulting privileged information $V_0$ is used for prediction by $gTr$ to generate $\hat{n}^{gTr}_0$. Subsequently, in each time frame $t$, a BB trial is run with rough estimate $\hat{n}^{gTr}_{t-1}$, and the privileged vector $V_t$ is fed to the $gTr$ network. The feature vector $(V_t, n_{t-1}^{truth})$ and the target $n_{t}^{truth}$ are stored in the teacher $Tr$ training data. The genie $gTr$ is  trained on the current data point with the loss $\mathcal{L}^{gTr}_t$ given by:
\begin{equation}
\mathcal{L}^{gTr}_t= (\hat{n}^{gTr}_t - n^{truth}_{t})^2. 
\end{equation}

In \textsc{train\_teacher\_offline}, the generated $Tr$ training data is shuffled to avoid overfitting. The dataset is split into $train$ and $test$, and a new teacher network $Tr$ is trained on the dataset using the MSE loss given by:
\begin{equation}
\mathcal{L}^{Tr}_t =L(\hat{n}^{Tr}_t, n^{truth}_{t}) =(\hat{n}^{Tr}_t - n^{truth}_{t})^2.
\end{equation}

\begin{algorithm}
	\caption{Teacher Training}\label{alg:teacher}
	\begin{algorithmic}[1]
		\Function{gen\_teacher\_training\_data}{}
		\State Randomly initialize $gTr$'s weights
        \State At $t=0$, conduct LoF trials to give the initial rough estimate $\hat{n}'_0$; then conduct a BB trial to generate $V_0$
		\State $\hat{n}^{gTr}_0 = g^{gTr}(V_0, \hat{n}'_{0})$
		\For{$t=1, \ldots, \text{num\_iters}$}
		\State	 Run BB trial with rough estimate $n'= \hat{n}_{t-1}^{gTr}$ to generate $V_t$	
		\State 	$\hat{n}^{gTr}_t = g^{gTr}(V_t, n^{truth}_{t-1})$
		\State Save $((V_t, n^{truth}_{t-1}), n^{truth}_{t})$ in $Tr$ training data
		\State Fit $g^{gTr}(.)$ to $((V_t, n^{truth}_{t-1}), {n}^{truth}_t )$ using loss $\mathcal{L}^{gTr}_t= (\hat{n}^{gTr}_t - n^{truth}_{t})^2$
		\EndFor
		\State \Return {$Tr$ training data} 
		\EndFunction
		
		\Function{train\_teacher\_offline}{$Tr$ training data}
		\State Randomly initialize teacher $Tr$
		\State Shuffle and split dataset into train and test
		\State Train $Tr$ with loss $\mathcal{L}^{Tr}_t = (\hat{n}^{Tr}_t - n^{truth}_{t})^2$
		\State {\Return $Tr$}
		\EndFunction

	\end{algorithmic}
\end{algorithm}
\begin{algorithm}
	\caption{Student Training}\label{alg:student}
	\begin{algorithmic}[1]
		\Function{gen\_student\_training\_data}{$\alpha$}
		\State Randomly initialize $gStu$'s weights
        \State At $t=0$, conduct LoF trials to give the initial rough estimate $\hat{n}'_0$; then conduct a BB trial to generate $v_0$
		\State $\hat{n}^{gStu}_0 = g^{gStu}(v_0, \hat{n}'_{0})$
		\For{$t=1, \ldots, \text{num\_iters}$}
		\State	 Run BB trial with rough estimate $n'= \hat{n}_{t-1}^{gStu}$	
		\State 	$\hat{n}^{gStu}_t = g^{gStu}(v_t, \hat{n}^{gStu}_{t-1})$
		\State Save the tuple $((v_t, \hat{n}^{gStu}_{t-1}), n^{truth}_{t})$ in $Stu$ training data; also save $V_t$ for $Tr$ prediction
		\State Fit $g^{gStu}(.)$ to $((v_t, \hat{n}^{gStu}_{t-1}), {n}^{truth}_t )$
		with loss $\mathcal{L}^{gStu}_t=\alpha(\hat{n}^{gStu}_t - n^{truth}_{t})^2 + (1-\alpha)(\hat{n}^{Tr}_t - n^{truth}_{t})^2$
		\EndFor
		\State \Return {$Stu$ training data} 
		\EndFunction
		\Function{train\_student\_offline}{$Tr$, $Stu$ training data, $\alpha$}
		\State Randomly initialize $Stu$'s weights; load pre-trained teacher $Tr$ 
		\State Shuffle and split $Stu$ training data into train and test
        \State For a particular BB trial with results $(v_t, V_t)$ conducted at time $t$, calculate the $Stu$ and $Tr$  predictions $\hat{n}^{Stu}_t = g^{Stu}(v_t, \hat{n}^{gStu}_{t-1})$ and $\hat{n}^{Tr}_t = g^{Tr}(V_t, n^{truth}_{t-1})$
		\State Train $Stu$ with loss  $\mathcal{L}^{Stu}_t= \alpha(\hat{n}^{Stu}_t - n^{truth}_{t})^2 + (1-\alpha)(\hat{n}^{Tr}_t - n^{truth}_{t})^2$
		\State {\Return $Stu$}
		\EndFunction
	\end{algorithmic}
\end{algorithm}

The student model does not see the privileged information (number of active nodes transmitting in each slot, true value of the number of active nodes in the previous time frame). It only sees the result of each slot (no transmission, success, or collision). Thus, the student has $(v_t, \hat{n}^{Stu}_{t-1})$ to calculate $\hat{n}^{Stu}_t$, which is given by:
\begin{equation}
\hat{n}^{Stu}_t = g^{Stu}(v_t, \hat{n}^{Stu}_{t-1}).
\end{equation}    
Instead of the standard regression MSE loss, distillation loss is used, which includes the loss between the teacher's prediction and the truth, and is given by:
\begin{equation}
\label{EQ:homogeneous:distillation:loss}
\mathcal{L}^{Stu}_t = (1-\alpha) L(\hat{n}^{Tr}_t , n^{truth}_{t})+ \alpha L(\hat{n}^{Stu}_t , n^{truth}_{t}),
\end{equation}
where $\alpha$ is the mixing ratio.

Similar to the teacher model's training, the results of the trials executed in different time frames are recorded offline, and a new model is trained on the recorded data with random shuffling. This concludes the training of the student model. The student training is explained in Algorithm \ref{alg:student}. In \textsc{gen\_student\_training\_data}, similar to the teacher training protocol, a genie student network $gStu$ is initialized. To initialize the protocol, LoF trials are conducted and the initial estimate $\hat{n}'_0$ is generated. $gStu$ uses the result of the BB trial of time frame $t$, viz., $v_t$, and the estimate of the previous trial $\hat{n}^{gStu}_{t-1}$ to compute $\hat{n}^{gStu}_{t}$. The feature vector for the student $(v_t, \hat{n}^{gStu}_{t-1})$, the target $n^{truth}_t$ and the feature vector for the teacher $(V_t, n^{truth}_{t-1})$ are stored in the $Stu$ training data. $gStu$ is then fit on the current sample with the distillation loss, which is a combination of the data loss and the teacher loss. 
In \textsc{train\_student\_offline}, a new student model $Stu$ is initialized and trained on the recorded data and with the pre-trained teacher $Tr$'s predictions. The general information $v_t$ and the privileged information $V_t$ are used for $Stu$ and $Tr$ inference, giving $\hat{n}^{Stu}_t$ and $\hat{n}^{Tr}_t$, respectively, as the $Stu$ and $Tr$ estimates. $Stu$'s weights are adjusted for minimizing the distillation loss $\mathcal{L}^{Stu}_t= \alpha(\hat{n}^{Stu}_t - n^{truth}_{t})^2 + (1-\alpha)(\hat{n}^{Tr}_t - n^{truth}_{t})^2$. Intuitively, a high $\alpha$ gives less importance to the $Tr$ loss and vice versa. 

\subsubsection{Other Algorithms for Comparison}
As a benchmark for comparison with the student model, in each time frame, the SRC$_s$ protocol (Algorithm \ref{alg:srcs}) is executed, with the number of time slots used being the same as the length of the BB trial used by the NN (student) model. There is an inherent disadvantage to the \srcs protocol since it does not use knowledge (e.g., estimate of the number of active nodes) from the previous time frame, unlike the NN method. To analyse whether the NN method estimates the number of active nodes better than an \srcs based algorithm that uses knowledge of an estimate of the number of active nodes in the previous time frame, we compare the former against the algorithm \emph{BB-Aware}, which is described in Algorithm \ref{alg:bb-aware}. BB-Aware uses \srcs in time frame $0$ to generate a rough estimate $\hat{n}^{BB-Aware}_0$. In each subsequent time frame $t$, BB-Aware conducts a BB trial of length $l_{BB-Aware}$ and with rough estimate $\hat{n}^{BB-Aware}_{t-1}$ (estimate of number of active nodes in the previous time frame) and computes the estimate $\hat{n}^{BB-Aware}_{t}$ by counting the number of empty slots in the trial (as in BB in Algorithm \ref{alg:bb}). 

For a fair comparison, each algorithm takes the same number of time slots to execute in each time frame. For example, if the total number of time slots in a time frame is to be $100$, then the NN method performs a BB trial of length $100$, the \srcs protocol performs $3$ LoF trials of length $8$ each, and a BB trial of length $76$, while the BB-Aware method performs a BB trial of length $100$.
 \begin{algorithm}
 	\caption{BB-Aware}
 	\label{alg:bb-aware}
 	\begin{algorithmic}[1]
 		\State In time frame $t=0$, execute \srcs to get $\hat{n}^{BB-Aware}_0$ 
 		\For{$t=1, \ldots, \text{num\_iters}$}
 		\State Conduct  a BB trial of length $l_{BB-Aware}$, with rough estimate $\hat{n}^{BB-Aware}_{t-1}$ 
            \State {\Return BB-Aware estimate $\hat{n}^{BB-Aware}_{t}$}
 		\EndFor
 	\end{algorithmic}
 \end{algorithm}

\subsection{Heterogeneous Network}
\label{SSC:algorithms:heterogeneous:network}
\subsubsection{Proposed Algorithm}
Consider the model and problem formulation described in Section \ref{SSC:heterogeneous:model:pf}. In this case, the coverage area of a base station contains $T$ types of nodes. The problem is to  estimate $\mathbf{n}^{truth}_t$, a $1 \times T$ vector. 

The approach followed is largely similar to that described in Section \ref{subsubsec:homogeneous_proposed}. Recall from Section \ref{subsubsec:homogeneous_proposed} that in the homogeneous case, a BB trial is conducted in each time frame; instead, in the heterogeneous case, in each time frame, 3-SS-BB (explained in Algorithm \ref{alg:3_ss_bb}) is conducted, and the results of all the slots are converted into a feature vector. 

\begin{figure}
	\centering
	\includegraphics[width=0.50\textwidth]{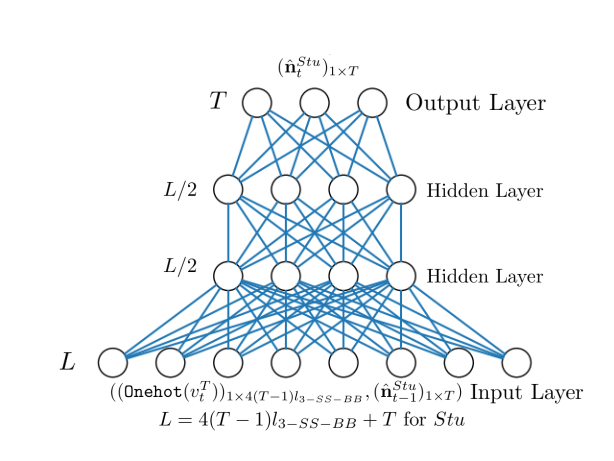}
	\caption{The figure shows the neural network architecture of a student ($Stu$) model used in the case of a heterogeneous network.}
	\label{fig:het_net}
\end{figure}

The architecture of the NN used is shown in Figure \ref{fig:het_net}. The input dense layer of length $L$ has relu activation, while the other two $L/2$ dense layers have sigmoid activation. The output layer of length $T$ has linear activation. A description of the activation functions is provided in \cite{dubey2022activation}.

\subsubsection{Training}
The teacher is trained on the feature vector $V^T_t$, which contains the number of nodes of each type participating in each of the $l$ blocks of 3-SS-BB, and $\mathbf{n}^{truth}_{t-1}$; note that the size of $\left( V^T_t, \mathbf{n}^{truth}_{t-1} \right)$ is  $(l+1)T$. The student is trained on $v^T_t$, which contains the result of each slot ($0, \alpha, \beta$ or $c$ (see Algorithm \ref{alg:3_ss_bb})) in each of the $l$ blocks of 3-SS-BB in one-hot encoding format, and $\mathbf{\hat{n}}^{Stu}_{t-1}$, which is the vector of estimates of the numbers of active nodes of the $T$ types produced by the student in time frame $t-1$; note that the size of $\left( v^T_t, \mathbf{\hat{n}}^{Stu}_{t-1} \right)$ is $4(T-1)l + T$. The training of the teacher and student proceed as per the procedures in Algorithms \ref{alg:teacher} and \ref{alg:student}, respectively, with the feature vectors being as in the heterogeneous case. 

\subsubsection{Other Algorithms for Comparison}
As a benchmark for comparison with the student model, in each time frame, SRC$_s$ is independently run $T$ times (henceforth referred to as T-SRC$_s$)-- once for each type of node. Similarly, the BB-Aware algorithm in Algorithm \ref{alg:bb-aware} is adapted to the heterogeneous case to give the algorithm T-BB-Aware, wherein in each time frame, BB-Aware is independently run $T$ times-- once for each type of node.

For a fair comparison, each algorithm takes the same number of time slots to execute in each time frame. In particular, the lengths of the BB trials in 3-SS-BB and T-\srcs are related as in the following equation:  
\begin{equation}
	\label{eq:tsrcs_trial_len}
	l_{3-SS-BB}(T-1) = (l_{SRC_s} + n_{LoF}l_{LoF})T.
\end{equation}
The length of a trial in 3-SS-BB is initially fixed and the length of a trial for \srcs ($l_{SRC_s}$) is computed using (\ref{eq:tsrcs_trial_len}) and rounded to the nearest integer.
 The lengths of the BB trials in 3-SS-BB and T-BB-Aware are related as in the following equation: 
\begin{equation}
    \label{eq:tbbaware_trial_len}
    l_{3-SS-BB}(T-1) = l_{BB-Aware}T.
\end{equation}
The length of a trial in 3-SS-BB is fixed and the length of the BB trials in T-BB-Aware,  $l_{BB-Aware}$, is computed using (\ref{eq:tbbaware_trial_len}) and rounded to the nearest integer. 

\section{Performance Evaluation}
\label{SC:simulations}
We describe the simulation setup in Section \ref{SSC:simulation:setup}. In Section \ref{SSC:simulations:homogeneous:network} (respectively, Section \ref{SSC:simulations:heterogeneous:network}), we provide our simulation results for the case of a homogeneous (respectively, heterogeneous) network. 

\subsection{Simulation Setup}
\label{SSC:simulation:setup}
\begin{figure}
	\centering
	\begin{subfigure}{0.45\textwidth}
		\centering
		\includegraphics[width=\textwidth]{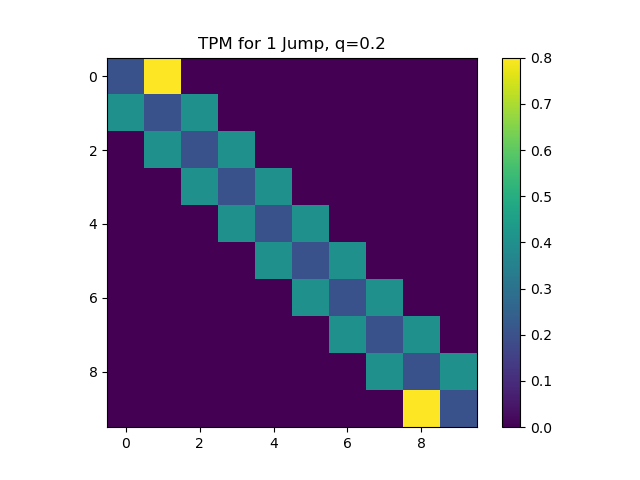}
		\caption{$P$ for $q=0.2$}
		\label{fig:tpm_j1}	
\end{subfigure}
	\hfill
	\begin{subfigure}{0.45\textwidth}
		\centering
		\includegraphics[width=\textwidth]{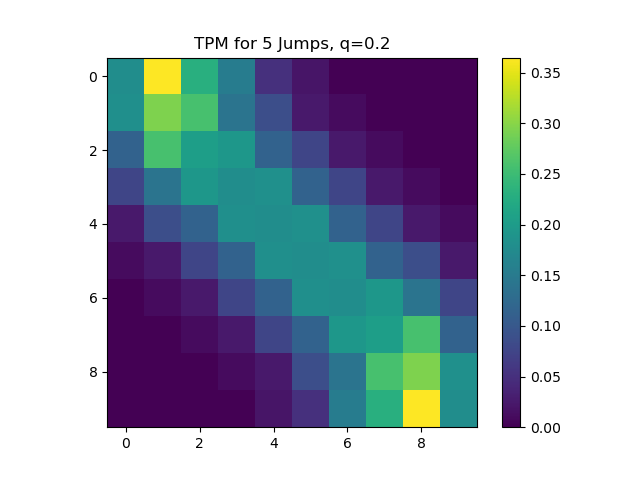}
		\caption{$P^5$ for $q=0.2$}
		\label{fig:tpm_j5}	
\end{subfigure}
	\caption{Figures (a) and (b) show the transition probability matrices $P$ and $P^5$, respectively, for $q = 0.2$. }
	\label{fig:tpm}
\end{figure}

The evolution of the number of active nodes of a given type over different time frames is modeled by a Discrete-Time Markov Chain (DTMC) with $N$ states $\{0, 1, \dots, N-1\}$, with a transition probability matrix (TPM) $P = [p_{i,j}]$, where the transition probabilities are given by the following equation:
\begin{equation}
\label{eq:tpm}
p_{i,j} = \begin{cases}
q, & \text{if } i=j,\\
1-q, & \text{if } i=0,j=1,\\
1-q, & \text{if } i=N-1,j=N-2,\\
1-p-q, & \text{if } i \neq 0, N-1 \text{ and } j=i-1, \\
p, & \text{if } i \neq 0, N-1 \text{ and } j=i+1.\\
\end{cases}
\end{equation}
We consider the case when $p=(1-q)/2$, which indicates that it is equally likely to go from a state $i$ to states $i+1$ and $i-1$. A visual representation of $P$ is shown in Figure \ref{fig:tpm_j1}. In order to model more sudden changes, we consider the $k$-step transition probability matrix ($P^k$), which allows changes in the number of active nodes from one time frame to the next one by more than $1$. An example is shown in Figure \ref{fig:tpm_j5}.

In the homogeneous case, let $n_t^{truth}$ be the number of active nodes in time frame $t$. The system evolves as in the following equation:
\begin{equation}
\label{eq:n_t_dtmc}
\mathbb{P}\lbrack n_t^{truth} = j \vert n_{t-1}^{truth} = i \rbrack = P^k[i, j],
\end{equation}
where $P^k[i, j]$ is the $(i,j)$'th element of the matrix $P^k$.  Also, in the heterogeneous case, the number of active nodes of each type evolves as in (\ref{eq:n_t_dtmc}), with the DTMCs for different types of nodes being independent. 

Throughout the simulations, for the homogeneous case, the maximum number of active nodes in a time frame is taken to be $64$ and for the heterogeneous case, the maximum number of active nodes of each type in a time frame is taken to be $\lfloor 192/T \rfloor$, where $T$ is the number of types.

\subsection{Homogeneous Network}
\label{SSC:simulations:homogeneous:network}
\begin{figure}
	\centering
	\includegraphics[width=0.85\linewidth]{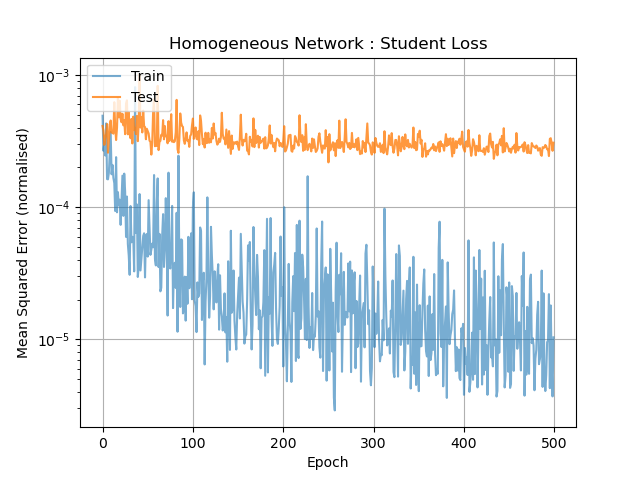}
	\caption{The figure shows the training of the  student using the teacher over epochs with $l_{BB}=100, \alpha=0.1, k=5$, a teacher dataset of $10^4$ time frames and a student dataset of $10^4$ time frames.}
	\label{fig:train_student_loss}
\end{figure}
Throughout the paper, an ``epoch'' refers to one complete pass through the entire training dataset during training. We plotted the evolution of the training and test loss over $2500$ epochs, when the student is trained using the teacher.
We observed that around $500$ epochs, the test loss increases significantly, while having insignificant variation in the training loss. The training of the student was thus stopped at that point.
Figure \ref{fig:train_student_loss} shows the evolution of the training and test loss over $500$ epochs, when the student is trained using the teacher.  The mixing ratio used in the training is $\alpha=0.1$. 

\begin{figure}
	\centering
	\includegraphics[width=0.85\linewidth]{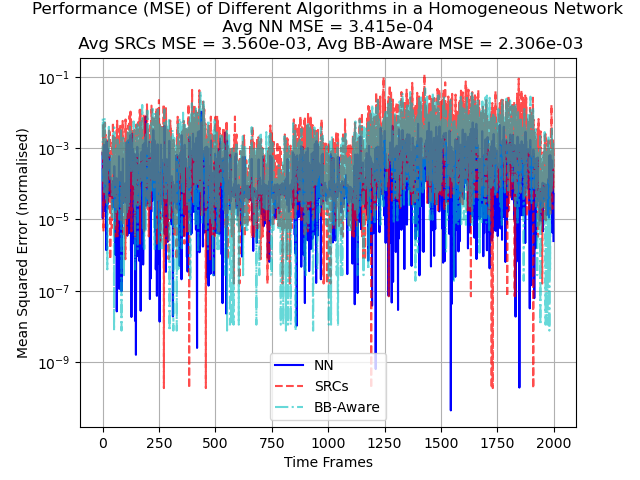}
	\caption{The figure shows the normalized mean squared errors (MSE) achieved by the NN, SRC$_s$, and BB-Aware methods, for a student network with $l_{BB}=100, \alpha=0.1, k=5, \text{num\_lof}=3, l_{lof} = 8$, a teacher dataset of $10^4$ time frames and a student dataset of $10^4$ time frames.}
	\label{fig:mse_default}
\end{figure}
Figure \ref{fig:mse_default} shows the normalized mean squared errors (MSE) in the active node cardinality estimates computed by  the trained student NN, the SRC$_s$ protocol, and the BB-Aware method, in different time frames in a homogeneous network. It is seen that the student NN achieves much lower normalized MSEs than both the other methods.

\begin{figure}
	\centering
	\includegraphics[width=0.85\linewidth]{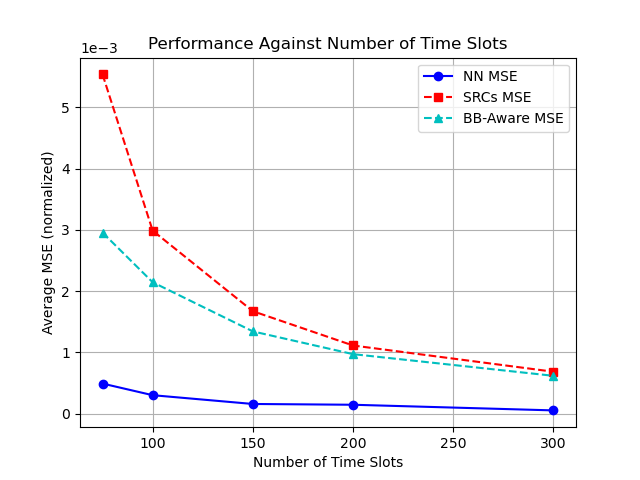}
	\caption{The figure shows the performance of different methods versus the number of time slots per time frame, with $\alpha=0.1, k=5, \text{num\_lof}=3, l_{lof} = 8$, a teacher dataset of $10^4$ time frames and a student dataset of $10^4$ time frames.}
	\label{fig:exp1}
\end{figure}
\textbf{Experiment 1 : Changing the length of the BB trial}\\
To study the variation of the error on changing the length of the BB trial, a different student network was trained for each length of trial. Each trained student network was then evaluated and the performance averaged over $20$ runs of $2000$ time frames each. The normalised MSE was then compared with those of the SRC$_s$ protocol and BB-Aware protocol. The result is shown in Figure \ref{fig:exp1}. It is seen that the errors of all the methods decrease with an increase in the length of trial, which is to be expected because a longer trial provides more information about the number of active nodes than a short trial, as there are less collisions. The NN performs better than \srcs and BB-Aware consistently. Hence, for achieving the same normalized MSE, a NN can make use of a shorter trial than the \srcs and BB-Aware protocols. \\
\textbf{Conclusion}: An increase in the length of the BB trial causes the errors of NN, \srcs and BB-Aware to decrease, with NN outperforming SRC$_s$ and BB-Aware for every length. 

\begin{figure}
	\centering
	\includegraphics[width=0.85\linewidth]{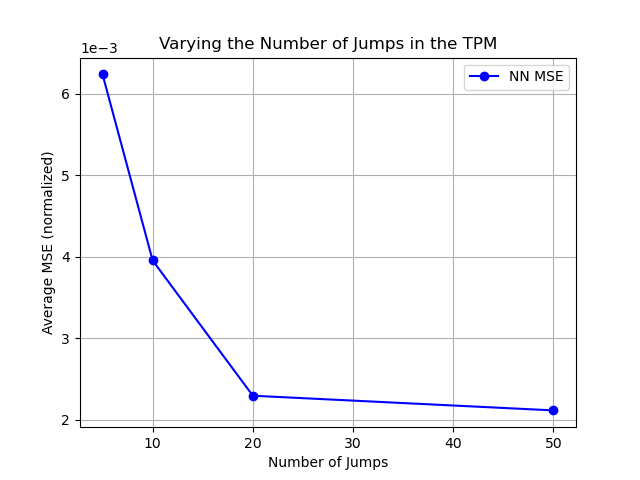}
	\caption{The figure shows the performance of the student network versus the number of jumps in the transition probability matrix, for a student network with $l_{BB}=100, \alpha=0.1$, a teacher dataset of $10^4$ time frames and a student dataset of $10^4$ time frames.}
	\label{fig:exp2}
\end{figure}
\begin{figure}
	\centering
	\includegraphics[width=0.85\linewidth]{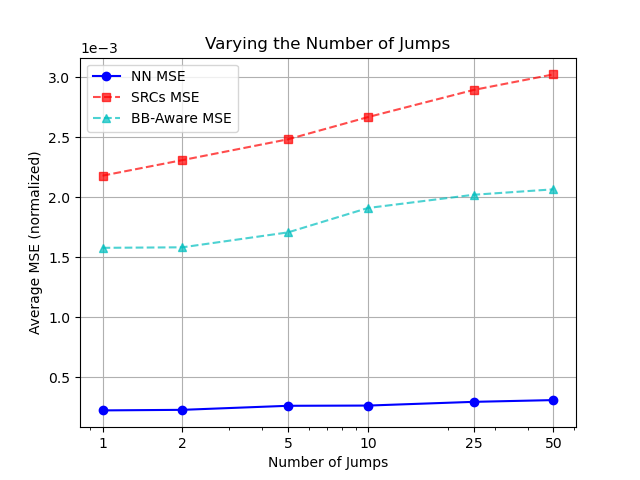}
	\caption{The figure shows the MSE achieved by different algorithms versus the number of jumps with $l_{BB}=100, \alpha=0.1, \text{num\_lof}=3, l_{lof} = 8$, a teacher dataset of $10^4$ time frames and a student dataset of $10^4$ time frames. The same student model trained on $5$ jumps is used for the estimation for every value of the number of jumps.}
	\label{fig:exp3}
\end{figure}
\textbf{Experiment 2 : Changing the number of jumps $k$}\\
To study how the methods perform when the system evolves faster or slower, the DTMC according to which the number of active nodes in a time frame evolves (see Section \ref{SSC:simulation:setup}) is changed by changing the number of jumps taken in one time frame. Specifically, for a DTMC with transition probability matrix $P^k$, by changing the number of jumps $k$, one can model a faster or slower changing time series. For each situation, a different student network is trained and the performance is evaluated by averaging over $20$  runs. Figure \ref{fig:exp2} shows that the normalized MSE achieved by the NN decreases with an increase in the number of jumps. For the same number of time frames that the student is trained on, a DTMC with higher $k$ offers more variation, and thus the NN is trained better and its achieved error decreases.\\
\textbf{Conclusion}: As the NN model is trained on more `adverse' scenarios when the number of jumps in the DTMC is higher (more outliers), the error decreases with an increase in the number of jumps. 

\textbf{Experiment 3 : Testing the same NN model with fast or slow time series}\\
If a single trained model is evaluated with different DTMCs (different values of $k$), the error does not vary much, as seen in Figure \ref{fig:exp3}. Note that in faster varying systems (higher $k$), the estimate from the previous time frame, which is used as the rough estimate for the BB trial in the current time frame, is more unreliable. The fact that despite this, the error achieved by the NN does not increase much in $k$ suggests that the student network has learned to map the results of the current trial to the number of active nodes well, rather than relying heavily on the estimate of the previous time frame.  BB-Aware performs worse under a faster time series (higher $k$), which is expected since it uses its estimate from the previous time frame as the rough estimate for the BB trial of the current time frame and  the correlation between the true values of the number of active nodes in the previous time frame and in the current time frame decreases as $k$ increases. Figure \ref{fig:exp3} also shows that for all values of $k$, NN significantly outperforms BB-Aware and SRC$_s$.  \\
\textbf{Conclusion}: The NN model learns the dependence between the $v_t$ vector and the target $n^{truth}_t$, and does not simply repeat $\hat{n}^{Stu}_{t-1}$; also, for all values of $k$, it significantly outperforms BB-Aware and SRC$_s$.

\begin{figure}
	\centering
	\includegraphics[width=0.85\linewidth]{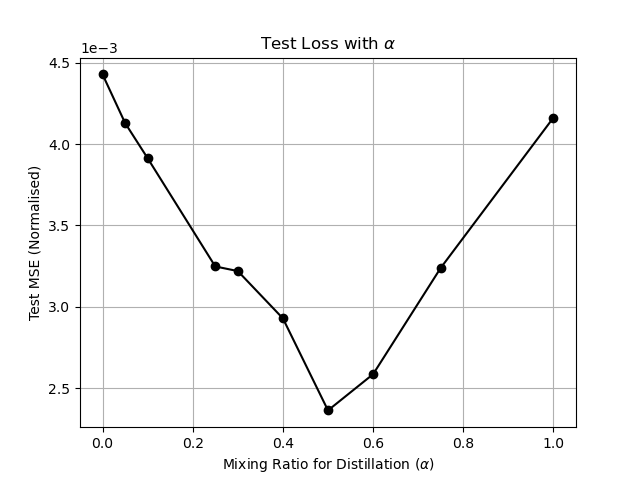}
	\caption{The figure shows the variation of the test loss with the mixing ratio $\alpha$ for a student network with $l_{BB}=100, k=5$, a teacher dataset of $10^4$ time frames and a student dataset of $10^4$ time frames.}
	\label{fig:test_alpha}
\end{figure}

\textbf{Experiment 4 : Varying the mixing ratio $\alpha$} \\
Next, we study the dependence of the test loss on the mixing ratio $\alpha$ (see (\ref{EQ:homogeneous:distillation:loss})). It can be seen from Figure \ref{fig:test_alpha} that the test loss decreases when $\alpha$ is  increased  up to a certain point, then increases again as $\alpha$ approaches  $1$. Figure \ref{fig:test_alpha} shows that distillation offers a significant benefit over blind training the student ($\alpha=1$). Also, the figure shows that the best result (lowest test loss) is achieved for a value of $\alpha$ around $0.5$. \\
\textbf{Conclusion}: Distillation offers a significant benefit over regular NN model training.

\subsection{Heterogeneous Network}
\label{SSC:simulations:heterogeneous:network}
\begin{figure}
	\centering
	\includegraphics[width=0.85\linewidth]{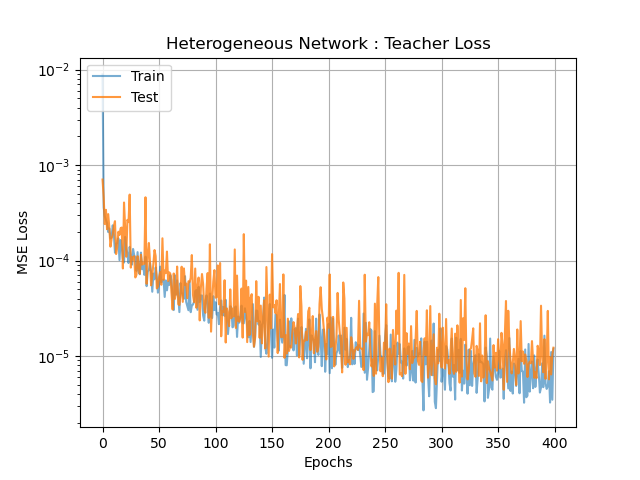}
	\caption{The figure shows the training of the teacher in the heterogeneous network case with $l_{3-SS-BB}=100, T=3, k=5$, and a teacher dataset of $10^4$ time frames.}
	\label{fig:het_teacher_loss}
\end{figure}

\begin{figure}
	\centering
	\includegraphics[width=0.85\linewidth]{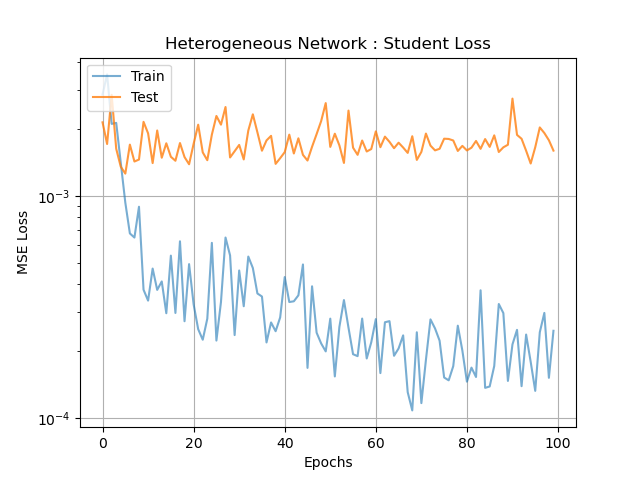}
	\caption{The figure shows the training of the student in the heterogeneous network case, with $l_{3-SS-BB}=100, T=3, \alpha=0.1, k=5$, a teacher dataset of $10^4$ time frames and a student dataset of $10^4$ time frames. }
	\label{fig:het_student_loss}
\end{figure}

Figure \ref{fig:het_teacher_loss} shows the training curves for offline training of a teacher model. It is seen that the test loss and training loss both decrease and settle quickly, indicating a relatively simple function to model. In contrast, Figure \ref{fig:het_student_loss} shows the training curves for offline training of the corresponding student model. As the student model is larger than the teacher model, it is slower to train. The function mapping the student input to the target is also significantly complex. This leads to the model overfitting to the training data, and the test loss remains roughly constant, while the training loss decreases. 

\begin{figure}
	\centering
	\includegraphics[width=0.85\linewidth]{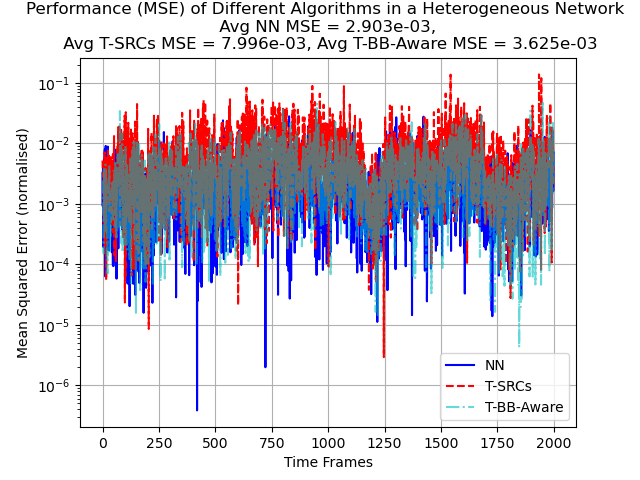}
	\caption{ The figure shows the normalized mean squared errors (MSE) achieved by the NN, T-SRC$_s$, and T-BB-Aware methods, with $l_{3-SS-BB}=100, T=3, \alpha=0.1, k=5, \text{num\_lof}=3, l_{lof} = 8$, a teacher dataset of $10^4$ time frames and a student dataset of $10^4$ time frames, and an evaluation run of 2000 time frames. }
	\label{fig:het_student_eval}
\end{figure}

After the training is complete, the trained student model is deployed in an online scenario. Figure \ref{fig:het_student_eval} shows the performances of the student model, T-SRC$_s$ and T-BB-Aware methods versus the time frame number for $1000$ time frames. It can be seen that the NN model significantly outperforms T-SRC$_s$ as well as T-BB-Aware.

\begin{figure}
	\centering
	\includegraphics[width=0.85\linewidth]{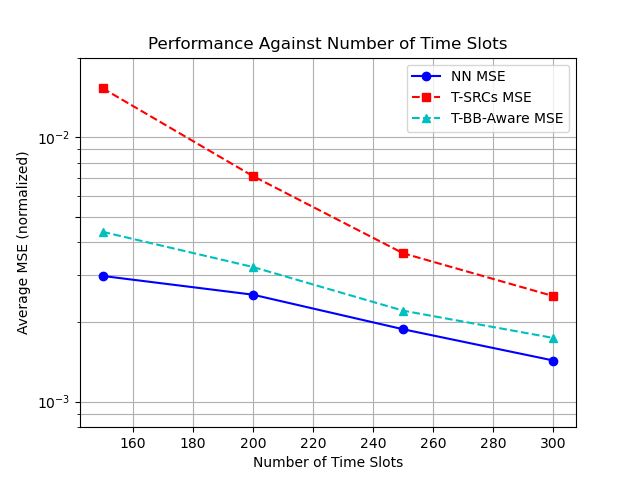}
	\caption{The figure shows the average MSE achieved by different algorithms versus the number of time slots taken by each algorithm in a time frame, with $T=3, \alpha=0.1, k=5, \text{num\_lof}=3, l_{lof} = 8$, a teacher dataset of $2 \times 10^4$ time frames and a student dataset of $2\times 10^4$ time frames.}
	\label{fig:het_exp1}
\end{figure}

\textbf{Experiment 1 : Changing the length of the trial}\\
The length of the trial $(l_{3-SS-BB})$ is changed, and a different student model is trained on the generated data for each value of $l_{3-SS-BB}$.  The other two methods, viz., T-\srcs and T-BB-Aware, are also configured to use the same total number of time slots per time frame to produce estimates as the NN method. As expected, in Figure \ref{fig:het_exp1}, the error decreases for the NN,  T-SRC$_s$, and T-BB-Aware methods as the number of time slots increases; this is because the number of collisions decreases under each method. The NN method significantly outperforms both the T-SRC$_s$ and T-BB-Aware methods, which indicates that to achieve the same average error, the NN method can deliver with a shorter trial than both T-\srcs and T-BB-Aware. T-BB-Aware outperforms T-SRC$_s$, as it uses a longer trial and uses information (estimates of numbers of active nodes of different types) from the previous trial. \\
\textbf{Conclusion}: The NN model requires a far shorter trial than T-\srcs and T-BB-Aware to offer a specified average MSE.

\begin{figure}
	\centering
	\includegraphics[width=0.85\linewidth]{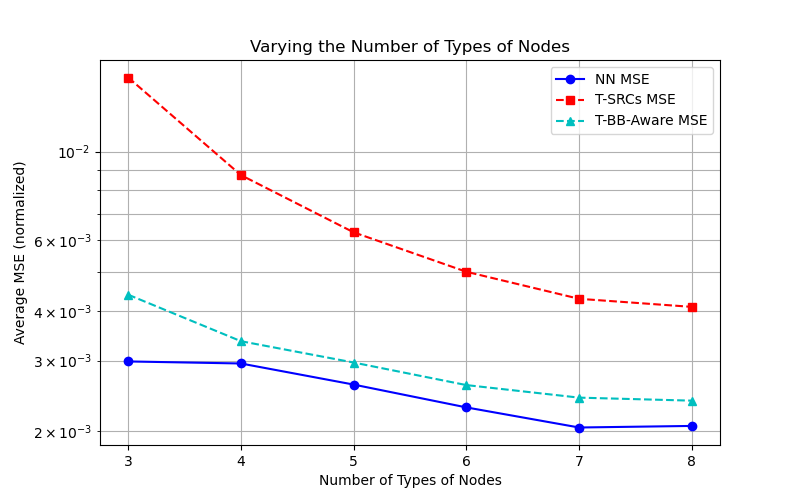}
	\caption{The figure shows the average MSE achieved by different algorithms versus the number of types of nodes ($T$) with $l_{3-SS-BB}=75, \alpha=0.1, k=5, \text{num\_lof}=3, l_{lof} = 8$, a teacher dataset of $2\times 10^4$ time frames and a student dataset of $2\times 10^4$ time frames.}
	\label{fig:het_exp2}
\end{figure}

\textbf{Experiment 2 : Changing the number of types of nodes}\\
Keeping the total number of nodes across all types present in the network the same (equal to $192$), the number of types of nodes ($T$) is now varied. The maximum number of active nodes of each type in a time frame is taken to be $\lfloor 192/T \rfloor$. The errors achieved by different methods are shown in Figure \ref{fig:het_exp2}. It is interesting to note that the NN method consistently gives low average MSE, even though the complexity of the mapping problem increases with $T$. Also, the figure shows that the NN method significantly outperforms both the T-SRC$_s$ and T-BB-Aware methods. \\
\textbf{Conclusion}: The NN method can adapt well to a higher number of types of nodes ($T$) while achieving significantly lower error than T-SRC$_s$ and T-BB-Aware for all values of $T$.

\section{Conclusions and Future Work}
\label{SC:conclusions:future:work}
We have proposed a novel methodology for node cardinality estimation in homogeneous as well as heterogeneous wireless networks, which uses the PFD technique and works using a neural network with a teacher-student model. Using extensive simulations, we have shown that the neural networks trained using PFD significantly outperform state-of-the-art node cardinality estimation algorithms. In particular, for a fixed number of time slots per time frame, the proposed PFD based algorithms  achieve much lower average normalised MSE than SRC$_s$ and $T$-SRC$_s$. Moreover, the proposed PFD based algorithms also outperform the  \srcs based BB-Aware and $T$-BB-Aware methods, which use information from the previous time frame and hence have longer BB trials, in homogeneous and heterogeneous wireless networks, respectively. Our work demonstrates that PFD is a promising approach for effectively solving the problem of node cardinality estimation in wireless networks. 

In this paper, we have assumed that the base station is stationary. A direction for future research is to extend the techniques proposed in this paper to the case where a mobile base station moves around, making multiple stops, for node cardinality estimation in a large region in which a homogeneous or heterogeneous wireless network is deployed. 